\documentclass[nohyper,12pt,letterpaper]{JHEP3}
\usepackage{epsfig}
\usepackage[latin1]{inputenc}
\usepackage{bbm,amsfonts}

\author{Marco Pirrone\\
Dipartimento di Fisica, Universit\`a di Milano--Bicocca and INFN, Sezione di Milano--Bicocca,
Piazza della Scienza 3, I-20126 Milano, Italy\\
\qquad\\
E-mail: \email{marco.pirrone@mib.infn.it}}

\abstract{We study the full bosonic spectrum around giant and dual giant graviton probes in exactly marginally deformed backgrounds. Considering supersymmetric and non--supersymmetric three--parameter 
deformations of $AdS_5\times S^5$, we perform a detailed analysis of small fluctuations for both the expanded D3--brane configurations. In particular, we enhance the scalar spectra of frequencies found in our previous paper hep-th/0609173 with the important contributions brought by the gauge field fluctuations. The giant graviton case exhibits a non--trivial coupling between scalar and vector modes driven by 
the deformation, whose resolution yields to a universal correction of the undeformed spectrum. On the other hand, dual giant vibrations turn out to be completely decoupled. From our results one can also easily read the gauge field fluctuations in the undeformed (dual) giant graviton scenario.}

\preprint{Bicocca--FT--08--1}

\title{Giants on Deformed Backgrounds Part II: The Gauge Field Fluctuations}


\csname @addtoreset\endcsname{equation}{section}

\def\bseq{\begin{subequation}}  
\def\eseq{\end{subequation}}
\def\bsea{\begin{subeqnarray}}  
\def\esea{\end{subeqnarray}}

\newcommand{\beq}{\begin{equation}}
\newcommand{\bea}{\begin{eqnarray}}
\newcommand{\eea}{\end{eqnarray}}
\newcommand{\eeq}{\end{equation}}

\newcommand {\non}{\nonumber}

\renewcommand{\d}{\delta}
\newcommand{\pa}{\partial}
\newcommand{\g}{\gamma}

\renewcommand{\l}{\lambda}

\def\hg{\hat{\gamma}}

\def\Mb{\kern 2pt\mathchoice
        {
         \vbox{\hrule width10pt height 0.4pt depth 0pt
         \kern 1.2pt\hbox{\kern -2pt$\displaystyle M$}}}
        {
         \vbox{\hrule width10pt height 0.4pt depth 0pt
         \kern 1.2pt\hbox{\kern -2pt$\textstyle M$}}}
        {
\vbox{\hrule width6pt height 0.4pt depth 0pt
         \kern 1.0pt\hbox{\kern -2pt$\scriptstyle M$}}}
        {
         \vbox{\hrule width5pt height 0.4pt depth 0pt
         \kern 0.8pt\hbox{\kern -2pt$\scriptscriptstyle M$}}}}

\def\Sb{\kern 2pt\mathchoice
        {
         \vbox{\hrule width6pt height 0.4pt depth 0pt
         \kern 1.2pt\hbox{\kern -2pt$\displaystyle S$}}}
        {
         \vbox{\hrule width6pt height 0.4pt depth 0pt
         \kern 1.2pt\hbox{\kern -2pt$\textstyle S$}}}
        {
         \vbox{\hrule width3.5pt height 0.4pt depth 0pt
         \kern 1.0pt\hbox{\kern -2pt$\scriptstyle S$}}}
        {
         \vbox{\hrule width3pt height 0.4pt depth 0pt
         \kern 0.8pt\hbox{\kern -2pt$\scriptscriptstyle S$}}}}

\def\Rb{\kern 2pt\mathchoice
        {
         \vbox{\hrule width5.5pt height 0.4pt depth 0pt
         \kern 1.2pt\hbox{\kern -2.5pt$\displaystyle R$}}}
        {
         \vbox{\hrule width5.5pt height 0.4pt depth 0pt
         \kern 1.2pt\hbox{\kern -2.5pt$\textstyle R$}}}
        {
         \vbox{\hrule width3.5pt height 0.4pt depth 0pt
         \kern 1.0pt\hbox{\kern -2.2pt$\scriptstyle R$}}}
        {
         \vbox{\hrule width3pt height 0.4pt depth 0pt
         \kern 0.8pt\hbox{\kern -2.2pt$\scriptscriptstyle R$}}}}

  \def\pp{{\mathchoice
      {
          \kern 1pt%
          \raise 1pt
          \vbox{\hrule width5pt height0.4pt depth0pt
            \kern -2pt
            \hbox{\kern 2.3pt
              \vrule width0.4pt height6pt depth0pt
              }
            \kern -2pt
            \hrule width5pt height0.4pt depth0pt}%
            \kern 1pt
       }
        {
          \kern 1pt%
          \raise 1pt
          \vbox{\hrule width4.3pt height0.4pt depth0pt
            \kern -1.8pt
            \hbox{\kern 1.95pt
              \vrule width0.4pt height5.4pt depth0pt
              }
            \kern -1.8pt
            \hrule width4.3pt height0.4pt depth0pt}%
            \kern 1pt
        }
        {
          \kern 0.5pt%
          \raise 1pt
          \vbox{\hrule width4.0pt height0.3pt depth0pt
            \kern -1.9pt  
            \hbox{\kern 1.85pt
              \vrule width0.3pt height5.7pt depth0pt
              }
            \kern -1.9pt
            \hrule width4.0pt height0.3pt depth0pt}%
            \kern 0.5pt
        }
        {
          \kern 0.5pt%
          \raise 1pt
          \vbox{\hrule width3.6pt height0.3pt depth0pt
            \kern -1.5pt
            \hbox{\kern 1.65pt
              \vrule width0.3pt height4.5pt depth0pt
              }
            \kern -1.5pt
            \hrule width3.6pt height0.3pt depth0pt}%
            \kern 0.5pt
        }
    }}

  \def\mm{{\mathchoice
               {
                 \kern 1pt
           \raise 1pt    \vbox{\hrule width5pt height0.4pt depth0pt
                  \kern 2pt
                  \hrule width5pt height0.4pt depth0pt}
                 \kern 1pt}
               {
                \kern 1pt
           \raise 1pt \vbox{\hrule width4.3pt height0.4pt depth0pt
                  \kern 1.8pt
                  \hrule width4.3pt height0.4pt depth0pt}
                 \kern 1pt}
               {
                \kern 0.5pt
           \raise 1pt
                \vbox{\hrule width4.0pt height0.3pt depth0pt
                  \kern 1.9pt
                  \hrule width4.0pt height0.3pt depth0pt}
                \kern 1pt}
               {
               \kern 0.5pt
         \raise 1pt  \vbox{\hrule width3.6pt height0.3pt depth0pt
                  \kern 1.5pt
                  \hrule width3.6pt height0.3pt depth0pt}
               \kern 0.5pt}
               }}

\def\pd{{\kern0.5pt
           + \kern-5.05pt \raise5.8pt\hbox{$\textstyle.$}\kern
0.5pt}}

\def\pmd{{\kern0.5pt
          \pm \kern-5.05pt
\raise6.3pt\hbox{$\textstyle.$}\kern1.5pt}}

\def\md{{\mathchoice
   {
      {{\kern 1pt - \kern-6.2pt \raise5pt\hbox{$\textstyle.$}\kern
1pt}}}
    {
      {{\kern 1pt - \kern-6.2pt \raise5pt\hbox{$\textstyle.$}\kern
1pt}}}
    {
      {\kern0.5pt - \kern-5.05pt
\raise3.4pt\hbox{$\textstyle.$}\kern0.5pt}}
    {
      {\kern0.5pt - \kern-5.05pt
\raise3.4pt\hbox{$\textstyle.$}\kern0.5pt}}}}

\begin{document}

\section{Introduction}

This is a natural sequel of our previous work~\cite{godb} in which we have studied giant and dual giant graviton configurations in the marginally deformed backgrounds proposed by Lunin--Maldacena (LM) and Frolov in~\cite{LM,F}.  

Giant gravitons were first introduced in~\cite{MGST} where expanded brane configurations in the $AdS_5\times S^5$ background with exactly the same quantum numbers of a point particle were found. In particular, they were described as stable D3--branes sitting at the center of $AdS_5$, wrapping an $S^3$ in the $S^5$ part of the geometry and traveling around an equator of the internal space. In~\cite{GMT,HHI} it was shown that also stable D3--branes blown up into AdS exist, the so--called dual giant gravitons. The main feature of both the configurations is to saturate a supersymmetric BPS bound for their energy, which turns out to be equal to their angular momentum in units of the radius of the background. 

In~\cite{godb}, the possible formation of (dual) giant gravitons was analyzed in the non--supersymmetric three--parameter $\gamma_i$--deformation of the $AdS_5\times S^5$ background~\cite{F}. A trivial translation of the results to the superconformal Lunin--Maldacena deformation~\cite{LM} was obtained by simply setting $\gamma_i=\gamma$.
We found stable states for extended D3--brane solutions grown into the deformed five--sphere $\tilde{S}^5$ and also inside the $AdS_5$ spacetime. The striking outcome was an identical scenario to the undeformed one: The dynamical branes were completely blind to the deformation. In particular, the supersymmetric $\gamma_i=\gamma$ case was not special as long as the procedure was totally independent of the specific value of the deformation parameters. We have also examined the scalar spectrum of small fluctuations around the ground state solutions, in order to study their stability. The main results of~\cite{godb} were that the deformed frequencies turned out to depend on the radius of the (dual) giant and the deformation entered into the spectrum bringing positive contributions. The conclusion was that the (dual) giant gravitons were perturbatively stable states. All the calculations were worked out with {\it a priori} vanishing gauge field fluctuations on the worldvolume of the branes. Is this a coherent procedure to have full control of the vibration spectrum?

Driven by recent results, the answer is now non--trivial. 
In fact, in~\cite{mim} the embedding of D7--branes was studied in LM--Frolov
backgrounds with the aim of finding the mesonic spectrum of the dual Yang--Mills theory 
with flavors, according to the gauge/gravity correspondence. In particular, we have considered a spacetime--filling D7--brane wrapped on a deformed three--sphere in the 
internal coordinates. We have found that for both the supersymmetric and
the non--supersymmetric deformations a static configuration existed which was 
completely independent of the specific values of the
deformation parameters $\gamma_i$. Moreover, by studying the fluctuations of the D7--brane, we have observed that the background deformation induces a non--trivial coupling between scalar and vector modes, crucial in the exact determination of the meson mass spectrum. With a field redefinition, we have simplified the equations of motion for the bosonic modes
and solved them analytically.

With this in mind, we now try to understand if the $U(1)$ worldvolume gauge fields play a similar role also for D3--brane giant and dual giant configurations.  Since the stable giant gravitons expanding in the deformed part of the geometry~\cite{godb} wrap the same cycle inside the internal space
as the D7--brane does~\cite{mim}, their bosonic fluctuations should encode the same features. On the other hand, we expect that nothing changes for dual giants and the scalar--vector coupling is absent because their worldvolume lies in AdS. 

In this respect, we first consider the equations of motion for the complete tower of bosonic Kaluza--Klein
modes arising from the compactification of a D3--brane giant on a deformed three--sphere.
As anticipated, we find a coupling between scalars and vectors which can be handled with a field redefinition similar to the one so useful in~\cite{mim}. 
Having performed this simplification, we obtain a universal and nice dependence on the deformation parameters in all the bosonic fluctuations, previously missed in~\cite{godb}. More precisely, all the undeformed frequencies are shifted by the same deformation--dependent quantity and the parameters $\g_{2,3}$, associated to TsT transformations
along the tori with a direction orthogonal to the probe brane, always appear in the complete spectrum. Exactly as in~\cite{mim} the parameter $\g_1$ 
related to the
deformation along the torus inside the D--brane worldvolume never enters the equations of motion for
quadratic fluctuations and does not affect the spectrum.
The corresponding states are classified according to their $SO(4)$ (the isomorphisms of the three--sphere)
and $U(1) \times U(1)$ quantum numbers $(\ell; m_2, m_3)$ and the universal shift depends on the deformation parameters through the linear combination $(\g_2 m_3 - \g_3 m_2)^2$. There exists a smooth limit to the undeformed case by sending $\g_i \to 0$ and this also gives us the behavior of the gauge field fluctuations of giant gravitons not included in the scalar analysis of~\cite{DJM}.  
The effect of the deformation is to break $SO(4)$ 
down to $U(1) \times U(1)$ and a Zeeman--like effect occurs. A suitable shift in $\ell$ gives $8(\ell+2)^2$ undeformed degenerate degrees of freedom but, because of the
$\g_i$--degeneracy breaking, they split among different frequency values. Closely following what has been done in~\cite{mim}, we find that
the splitting is different according to the choice $\g_2 \neq \g_3$ or $\g_2=\g_3$ (which
comprises the $\mathcal{N}=1$ supersymmetric deformation).

The dual giant graviton story is instead the same as in~\cite{godb}. We find no coupling between scalar and vector modes since the brane now expands in AdS. Moreover, the gauge fluctuations turn out to be completely independent of the deformation parameters so that also the undeformed case can be directly read from our computation.

The plan of the paper is as follows. After an introductory section on three--parameter deformations of the $AdS_5\times S^5$ and a review of the stable giant and dual giant graviton configurations on these backgrounds found in~\cite{godb}, in Section 3 we determine the exact and full bosonic spectrum of small fluctuations around the equilibrium giant solutions. We also discuss the properties of the spectrum and analyze in details the splitting of the frequency levels and 
the corresponding degeneracy. Finally, in Section 4 we 
consider the dual giant graviton case, whereas our conclusions follow in Section 5.

\section{Marginally deformed backgrounds and (dual) giant gravitons}\label{rev}
 
We start by giving the non--supersymmetric three--parameter deformation of the type IIB $AdS_5\times S^5$ background. This is realized by three TsT transformations (T duality--angle shift--T duality) along three tori inside $S^5$ and driven by three different real parameters $\g_i$~\cite{F}. Since the deformation is exactly marginal, the AdS factor remains unchanged. Let us write the metric of the deformed five--sphere $\tilde{S}^5$ (in string frame and setting for the moment $\alpha'=1$) using radial/toroidal coordinates $(\mu_i, \phi_i)$, $i=1,2,3$, with $\sum_i \mu_i^2 = 1$:
\beq\label{sdef}
ds^2_{\tilde{S}^5}=R^2\left(\frac{dr^2}{R^2-r^2}+\frac{r^2}{R^2}d\theta^2+G\sum_{i=1}^{3} \mu_i^2 d\phi_i^2\right) + R^2 G \mu_1^2 \mu_2^2 \mu_3^2 \left(\sum_{i=1}^{3}\hat{\gamma}_i d\phi_i\right)^2
\eeq
where $r\in[0, R]$, $\theta\in \left[0,\frac{\pi}{2}\right]$ and $\phi_i\in \left[0,2\pi\right]$. Recall that 
\beq
G^{-1}=1+\hat{\gamma}_1^2 \mu_2^2 \mu_3^2+\hat{\gamma}_2^2 \mu_1^2 \mu_3^2+\hat{\gamma}_3^2 \mu_1^2 \mu_2^2\,\, \qquad \qquad \hat{\gamma}_i=R^2 \gamma_i
\eeq
In what follows it will be convenient to parametrize $\mu_i$ coordinates via 
\beq
\mu_1^2=1-\frac{r^2}{R^2}\qquad\,\,\,\,\,\mu_2^2=\frac{r^2}{R^2}\cos^2{\theta}\qquad\,\,\,\,\,\mu_3^2=\frac{r^2}{R^2}\sin^2{\theta}
\eeq
Explicitly, the deformations involve the three tori $(\phi_1,\phi_2)$, $(\phi_1,\phi_3)$, 
$(\phi_2,\phi_3)$ and are parametrized by constants $\hg_3$, $\hg_2$ and $\hg_1$ respectively. In the case of real deformation parameters, in which we work, the axion field $C_0$ is vanishing while the non--constant dilaton is
\beq\label{dilaton}
e^{2\phi}=G
\eeq
with the undeformed one chosen to be zero. The background carries also a non--vanishing NS--NS two--form $B$ and R--R forms as well:
\bea\label{B}
B&=&R^2 G \left(\hat{\gamma}_3 \mu_1^2 \mu_2^2 d\phi_1 \wedge d\phi_2+\hat{\gamma}_1\mu_2^2\mu_3^2 d\phi_2\wedge d\phi_3+\hat{\gamma}_2\mu_3^2\mu_1^2 d\phi_3\wedge d\phi_1   \right)\non\\
C_2&=& -\frac{r^4}{R^2}  \sin{\theta}\cos{\theta} d\theta \wedge\sum_{i=1}^{3}\hat{\gamma}_i d\phi_i\non\\
C_4&=& \omega_4 + r^4 G \sin{\theta}\cos{\theta} d\theta \wedge d\phi_1\wedge d\phi_2\wedge d\phi_3
\eea
where $\omega_4$ is the four--form on the AdS part of the geometry. It is important to note that the combination 
\bea
C_4-C_2 \wedge B&=& \omega_4 + r^4 \sin{\theta}\cos{\theta} d\theta \wedge d\phi_1\wedge d\phi_2\wedge d\phi_3
\eea
is exactly the R--R four--form of the undeformed $AdS_5\times S^5$ spacetime~\cite{godb}. The supersymmetric Lunin--Maldacena background~\cite{LM} can be recovered by setting $\hg_1 = \hg_2 =\hg_3$.

\vskip 15pt
\noindent
We now follow~\cite{godb} and review the introduction of stable D3--brane giant and dual giant graviton configurations in these deformed backgrounds. The dynamics of bosonic degrees of freedom of a D3--brane is described by the action
\beq\label{dbiwz}
S=S_{DBI}+S_{WZ}=-T_3\int_{\Sigma_4} d^4\sigma\, e^{-\phi}\sqrt{-det(g+\mathcal{F})}+T_3\int_{\Sigma_4}P\left[\sum_n \,C_n\right] \,e^{\mathcal{F}}
\eeq
with $g$ denoting the pull--back of the ten--dimensional spacetime metric on the worldvolume $\Sigma_4$ of the brane parametrized by coordinates $\sigma^a$ (in what follows we choose a static gauge and latin labels $a,b,...$ stand for worldvolume components). The gauge potential $A$ enters the action through the $U(1)$ worldvolume gauge field strength $F$ in the modified field strength $\mathcal{F}=2 \pi \alpha' F-b$, where $b$ is the pull--back to the worldvolume of the target NS--NS two-form potential. $T_3$ is the D3--brane tension and in the Wess--Zumino term $P[...]$ denotes again the pull--back.
 
The first expanded solution we analyze is a D3--brane blown up inside the deformed five--sphere part of the geometry and sitting at the center of $AdS_5$: The giant graviton. In particular the brane wraps the $(\theta,\phi_2,\phi_3)$ directions, it has constant radius $r=r_0$ and orbits the $\tilde{S}^5$ in the $\phi_1$ direction with a constant angular velocity $\omega_0$. Writing the $AdS_5$ metric as
\beq\label{old}
ds^2_{AdS_5}=-(1+\frac{l^2}{R^2})dt^2+\frac{dl^2}{1+\frac{l^2}{R^2}}+l^2\left[d\alpha_1^2+\sin^2{\alpha_1}\left(d\alpha_2^2+\sin^2{\alpha_2}d\alpha_3^2\right)\right]
\eeq
our configuration is given by
\beq\label{prima}
r=r_0\,\qquad \phi_1=\omega_0 t\,\qquad l=\alpha_1=\alpha_2=\alpha_3=0\,\qquad F=0
\eeq
\noindent
which, after integration on the spatial coordinates of the worldvolume in (\ref{dbiwz}), leads to the effective Lagrangian
\beq\label{lgi}
L=-N \frac{r_0^3}{R^4}\,\left[\sqrt{1-(R^2-r_0^2)\, \dot{\phi_1}^2}- r_0\,\dot{\phi}_1\right]
\eeq
and to the corresponding Hamiltonian
\beq\label{ham}
H=\dot{\phi_1} J-L=\frac{N}{R^4}\,\sqrt{r_0^6+\frac{(R^4 J/N - r_0^4)^2}{R^2-r_0^2}}
\eeq
where $J$ is the conserved conjugate momentum to $\phi_1$ (see~\cite{godb} for more details). For fixed $J$, we have the interesting minimum of (\ref{ham}) at $r_0=R\sqrt{\frac{J}{N}}$, where the energy is $E=\frac{J}{R}$ and $\omega_0=\dot{\phi_1}=\frac{1}{R}$. We stress that this expanded D3--brane configuration has the same quantum numbers as the point--like graviton even in the deformed $AdS_5\times\tilde{S}^5$ background~\cite{godb} and the classical configuration is completely blind to the deformation parameters $\hg_i$.

On the other hand, there exists also the possibility of a dual giant graviton equilibrium solution where the D3--brane is wrapped on the three--sphere $(\alpha_1,\alpha_2,\alpha_3)$ contained in the $AdS_5$ part of the geometry.
The brane now has constant radius $l_0$ and again orbits rigidly in the $\phi_1$ direction of the $\tilde{S}^5$:
\beq\label{ansadu}
l=l_0\,\qquad \phi_1=\omega_0 t\,\qquad r=\phi_2=\phi_3=0\,\qquad \theta=\frac{\pi}{4}\,\qquad F=0
\eeq
\noindent
The effective Lagrangian and Hamiltonian for this configuration are
\bea
L &=&-N \frac{l_0^3}{R^4}\,\left[\sqrt{1+\frac{l_0^2}{R^2}-R^2\,\dot{\phi_1}^2}-\frac{l_0}{R}\right]\non\\
H&=&\frac{N}{R^4}\,\left[\sqrt{\left(1+\frac{l_0^2}{R^2}\right)\,\left(l_0^6+R^6\,\frac{J^2}{N^2}\right)}-\frac{l_0^4}{R}\right]
\eea
again introducing the conjugate momentum $J$ to $\phi_1$. The Hamiltonian $H$, as a function of $l_0$, has an expanded minimum located at $l_0=R\sqrt{\frac{J}{N}}$; the ground state energy is $E=\frac{J}{R}$ and $\omega_0=\dot{\phi_1}=\frac{1}{R}$, matching the results of the giant graviton case.  

As extensively stressed in~\cite{godb}, this is the same situation known from the standard undeformed $AdS_5\times S^5$ background. In fact, even for the deformed $AdS_5\times \tilde{S}^5$, there are three potential configurations to describe a graviton carrying angular momentum $J$: The point--like graviton, the giant graviton consisting of a D3--brane lying in the deformed five--sphere and the dual giant graviton consisting of a spherical three--dimensional brane which expands into the AdS space. The precise values of the deformation parameters $\hg_i$ never enter the calculation so the non--supersymmetric case behaves exactly as the supersymmetric $\hg_i=\hg$ one.

These facts represent strong hints on the stability under the perturbation of the equilibrium configurations found. 
In order to verify this expectation we have also studied the spectrum of small scalar fluctuations around the (dual) giant graviton solutions. The deformed frequencies turned out to depend on deformation parameters but their positive contributions
showed that the expanded gravitons were perturbatively stable states.

\section{The complete bosonic spectrum around giant gravitons}\label{vib}

Our main purpose is to investigate again oscillations of the D3--brane probes just described in these deformed backgrounds with the aim of studying the complete bosonic spectrum. So we reanalyze the set of small fluctuations about the expanded giant solutions together with their worldvolume gauge field fluctuations turned on.

Generic vibrations of a brane around its ground state can be described by  
\beq
X=X_0+\varepsilon \delta X(\sigma^a)
\eeq
and by a non--trivial flux $\varepsilon\,F = \varepsilon \,dA$. With $X$ we mean one of the spacetime coordinates, $X_0$ denotes the solution of one of the unperturbed equilibrium configurations of the previous section, the fluctuation $\delta X(\sigma^a)$ is a function of the worldvolume coordinates $\sigma^a$ and $\varepsilon\equiv 2\pi\alpha'$ is viewed as a small perturbation parameter. 

The action (\ref{dbiwz}) consists of the Dirac--Born--Infeld term
\beq\label{lag}
\mathcal{L}_{DBI}=-T_3\frac{1}{\sqrt{G}}\sqrt{-det(g-b+\varepsilon\,F)}
\eeq
where we have written the dilaton field as in (\ref{dilaton}), and of the Wess--Zumino one which reads
\beq\label{swz}
\mathcal{L}_{WZ}=T_3\Big\{P\left[C_4-C_2\wedge B\right]+\varepsilon P\left[C_2\right]\wedge F \Big\}
\eeq
We consider both of the terms expanded up to the quadratic order in $\varepsilon$, giving
\bea\label{expla}
S&=&\int_{\Sigma_4} d^4\sigma\left[\left(\mathcal{L}_{DBI}^{(0)}+\mathcal{L}_{WZ}^{(0)}\right)+\varepsilon\,\left(\mathcal{L}_{DBI}^{(1)}+\mathcal{L}_{WZ}^{(1)}\right)+\varepsilon^2\,\left(\mathcal{L}_{DBI}^{(2)}+\mathcal{L}_{WZ}^{(2)}\right)\right]\non\\
&\equiv&\int_{\Sigma_4} d^4\sigma\left(\mathcal{L}^{(0)}+\varepsilon\,\mathcal{L}^{(1)}+\varepsilon^2\,\mathcal{L}^{(2)}\right)
\eea

In this section we only focus on the giant graviton case. Instead of (\ref{old}), to study the fluctuation spectrum it is convenient to use the following parametrization of the $AdS_5$ part of the metric~\cite{DJM,godb}
\beq
ds^2_{AdS_5}=-\left(1+\sum_{k=1}^{4}v_k^2\right)dt^2+R^2\left(\delta_{ij}+\frac{v_i v_j}{1+\sum_{k=1}^{4}v_k^2}\right)dv_i dv_j
\eeq
and consider the modified ansatz
\beq
r=r_0+\varepsilon\,\rho(\sigma^a)\,\qquad \phi_1=\omega_0 t+\varepsilon\,\varphi(\sigma^a)\,\qquad v_k=\varepsilon\,\chi_k(\sigma^a)\,\qquad F=dA
\eeq
with respect to the D3--brane configuration given in (\ref{prima}), which now reads 
\beq\label{soluh}
r=r_0\equiv R \sqrt{\frac{J}{N}}\,\qquad \phi_1=\omega_0 t\equiv\frac{1}{R}\, t\,\qquad v_k=0\,\qquad F=0
\eeq
Recall that $F$ is always dressed with $\varepsilon$, $J$ is the fixed angular momentum and we have $\sigma^a=(t,\theta,\phi_2,\phi_3)$. To ease the presentation of our results we introduce the undeformed metric $\mathcal{G}_{ab}$ with non--vanishing entries
\beq\label{unmat}
\mathcal{G}_{tt}=-\frac{1}{R^2}\qquad \mathcal{G}_{ij}=\tilde{g}_{ij}
\eeq
where $\tilde{g}_{ij}=\mbox{diag}(1, c_\theta^2, s_\theta^2)$ is the diagonal metric on the unit three--sphere $(\theta,\phi_2,\phi_3)$ with determinant $\tilde{g} = \sin^2{\theta} \cos^2{\theta}$, and the fixed vector 
\beq
q^a = \hat{\g}_2 \d^a_{\phi_3} - \hat{\g}_3 \d^a_{\phi_2}
\label{vector}
\eeq
Obviously the $\mathcal{L}^{(0)}$ term in (\ref{expla}) gives a zeroth order Lagrangian density related to (\ref{lgi}), while we can write the first as well the second order corrections in the following compact forms
\bea\label{L2}
\mathcal{L}^{(1)} &=& T_3 r_0^2 \sqrt{\tilde{g}}\left[ R^2 \partial_t \varphi+\hat{\gamma}_2 F_{t \phi_3}-\hat{\gamma}_3 F_{t \phi_2}\right]
\non \\
\mathcal{L}^{(2)}&=&
-T_3 r_0^2 \sqrt{\tilde{g}} \left[ \frac{R}{2}\,\mathcal{C}^{ab}\partial_a \chi_k \partial_b 
\chi_k +  \frac{R}{2(R^2-r_0^2)}\,\mathcal{C}^{ab}\partial_a \rho \partial_b \rho + \frac{R(R^2-r_0^2)}{2 r_0^2}\,\mathcal{G}^{ab}\partial_a \varphi\partial_b \varphi \right.\non\\
&~&+\left. \frac{1}{4 R r_0^2}F_{ab}F^{ab}+ \frac{(R^2-r_0^2)}{R r_0^2}\,q^c F_{ac}\mathcal{G}^{ab} \partial_b \varphi -\frac{2 R^2}{r_0}\left(\partial_t \varphi + \frac{1}{R^2} q^c F_{t c} \right) \rho + \frac{R}{2}\,\chi_k^2\right]\non\\
\eea
where 
\beq
\mathcal{C}^{ab}\equiv\mathcal{G}^{ab}+\frac{(R^2-r_0^2)}{R^2}\,q^a q^b
\eeq
$F^{ab} \equiv \mathcal{C}^{ac}\mathcal{C}^{bd}F_{cd}$ and the sum over $k$ is understood.

The first order Lagrangian $\mathcal{L}^{(1)}$ is a total derivative since (\ref{soluh}) is the right solution which really minimizes the action~\cite{godb} even with the inclusion of gauge fluctuations. 

 In writing $\mathcal{L}^{(2)}$ some terms are integrated by parts and it is interesting to note that only the deformation
parameters $\hat{\gamma}_2$ and $\hat{\gamma}_3$, hidden in the vector (\ref{vector}), appear. In fact, at this order 
the dependence on $\hat{\gamma}_1$, associated to the torus inside the D3 worldvolume, 
completely cancels in a very similar fashion to~\cite{mim}. Here, we have also used the Bianchi identity $\partial_t F_{\phi_2\phi_3}+\partial_{\phi_2} F_{\phi_3 t}+\partial_{\phi_3} F_{t \phi_2}=0$ which eliminates a $\hg_1$--dependent contribution.  

We now determine the equations of motion coming from the quadratic Lagrangian $\mathcal{L}^{(2)}$ in (\ref{L2}). Avoiding the extremal cases $r_0=0$ and $r_0=R$, for the $\chi_k$, $\rho$ and $\varphi$ scalars we find, respectively
\beq\label{eomchi}
\partial_a\left(\sqrt{\tilde{g}}\,\mathcal{C}^{ab}\,\partial_b\chi_k \right)-\sqrt{\tilde{g}}\,\chi_k=0
\eeq
\beq\label{eomrho}
\partial_a\left(\sqrt{\tilde{g}}\,\mathcal{C}^{ab}\,\partial_b\rho \right)+\frac{2 (R^2-r_0^2)}{R r_0}\, \sqrt{\tilde{g}}\,\left(R^2 \partial_t \varphi + q^b F_{t b}\right)=0
\eeq
\beq\label{eomphi}
\partial_a\left[\sqrt{\tilde{g}}\,\mathcal{G}^{ab}\,\left(\partial_b\varphi+\frac{1}{R^2}\,q^c F_{bc}\right)\right]-\frac{2 R r_0}{(R^2-r_0^2)}\, \sqrt{\tilde{g}}\,\partial_t \rho=0
\eeq
whereas, using (\ref{eomphi}) the equations of motion for the gauge fields take the form
\bea\label{eoma}
\partial_a\left(\sqrt{\tilde{g}}\,\mathcal{G}^{ac}\mathcal{G}^{bd}\,F_{cd}\right)
-(R^2-r_0^2)\,\sqrt{\tilde{g}}\, q^d \partial_d \,
\left[ \mathcal{G}^{bc}\,\left( \partial_c \varphi + \frac{1}{R^2} q^f
F_{c f}\right)-\frac{2 R r_0 \delta^b_t}{(R^2-r_0^2)}\,\rho\right]=0\non\\
\eea
which come into two distinct classes,
according to $b=t$ or
$b=i\equiv \{\theta,\,\phi_2,\,\phi_3\}$. 

The scalar fluctuations $\chi_k$ decouple from the rest. On the other hand, $\rho$ and $\varphi$
interact non--trivially among each other as in the undeformed case but they also couple with the worldvolume gauge fields through terms proportional to
the deformation parameters. This situation is intimately related to the resolution of the mesonic spectrum performed in~\cite{mim} which we closely follow from now on.

Before solving the equations of motion (\ref{eomchi})--(\ref{eoma}) 
for scalar and vector modes, let us write the abelian flux in terms of its potential one--form,
$F_{ab} = \partial_a A_b - \partial_b A_a$, and choose
the gauge $A_t=0$. Moreover, in order to simplify the equations we introduce the special operators
\beq\label{ogamma}
\mathcal{O}_{\hg} \equiv -R^2 \partial_t^2 + \nabla_i \nabla^i + \left(1-\frac{r_0^2}{R^2}\right)(\hat{\gamma}_2\partial_3-\hat{\gamma}_3\partial_2)^2\,\qquad\,\mathcal{O}_0 \equiv \mathcal{O}_{\hg}|_{\hat{\gamma}_2 = 
\hat{\gamma}_3 =0}
\eeq
where we write $\pa_2 \equiv \pa_{\phi_2}$ and $\pa_3 \equiv \pa_{\phi_3}$ for concision. Of course, by $\nabla_i$ we mean covariant derivatives on the unit three--sphere.

Using the general identity 
$\frac{1}{\sqrt{\tilde{g}}} \partial_i (\sqrt{\tilde{g}}\partial^i s) = \nabla_i \nabla^i s$ 
valid for any scalar $s$, equation (\ref{eomchi}) for the $\chi_k$ modes then takes the compact form 
\beq\label{ceom}
\left(\mathcal{O}_{\hg}-1\right)\,\chi_k=0
\eeq
whereas defining
\beq\label{Phi}
\Phi\equiv \varphi + \frac{1}{R^2} q^a A_a = 
\varphi+\frac{1}{R^2}(\hat{\gamma}_2 A_{\phi_3}-\hat{\gamma}_3 A_{\phi_2}) 
\eeq
equations (\ref{eomrho}) and (\ref{eomphi}) can be rewritten as
\beq\label{reom}
\mathcal{O}_{\hg}\,\rho + \frac{2 R (R^2-r_0^2)}{r_0}\,\partial_t \Phi=0
\eeq
\beq\label{peom}
\mathcal{O}_{0}\,\Phi-\frac{1}{R^2}(\hat{\gamma}_2\partial_3-\hat{\gamma}_3\partial_2)\,\nabla^i A_i- \frac{2 R r_0}{(R^2-r_0^2)}\,\partial_t \rho=0
\eeq

If $b=t$, equation (\ref{eoma}) for the vector modes becomes
\beq\label{ti}
\partial_t \left[\nabla^i A_i + (R^2-r_0^2) (\hat{\gamma}_2\partial_3-\hat{\gamma}_3\partial_2)\,\Phi\right] + \frac{2 r_0}{R}\,(\hat{\gamma}_2\partial_3-\hat{\gamma}_3\partial_2)\,\rho=0
\eeq
with $\Phi$ defined in (\ref{Phi}). On the other hand, if $b$ runs on the coordinates that parametrize the unit three--sphere, from (\ref{eoma}) we obtain
\bea\label{ii}
&&\mathcal{O}_{\hg}\,A_j-\nabla^i \nabla_j A_i-(R^2-r_0^2)(\hat{\gamma}_2\partial_3-\hat{\gamma}_3\partial_2)\,
\partial_j\Phi =0
\eea
where we have used $\frac{1}{\sqrt{\tilde{g}}} \pa_i (\sqrt{\tilde{g}} F^{ij}) = \nabla_i F^{ij}= 
\nabla_i \nabla^i A^j - \nabla_i \nabla^j A^i$. 

It is convenient to search for solutions expanded as plane--waves in $t$ (with frequency denoted by $\omega$)  and as spherical harmonics on $S^3$. In particular, it is natural to expand $\chi_k$, $\rho$ and $\Phi$ in scalar spherical harmonics and the gauge fields in vector ones (see for example the approach in~\cite{my}).

The scalar spherical harmonics are a complete set of functions $\mathcal{Y}_\ell^{m_2,m_3}$ 
in the $\left(\frac{\ell}{2},\frac{\ell}{2}\right)$ representation of $SO(4)$ and  
with definite $U(1)\times U(1)$ quantum numbers $(m_2,m_3)$ satisfying $|m_2\pm m_3| = \ell - 2k$, $\ell,k =0,1,\dots$. For fixed $\ell$ the degeneracy
is $(\ell+1)^2$.
Their defining equations are
\bea\label{Yprop}
\nabla_i \nabla^i\,\mathcal{Y}_\ell^{m_2,m_3}&=&
-\ell(\ell+2) \, \mathcal{Y}_\ell^{m_2,m_3}\non\\
\partial_{2,3}\,\mathcal{Y}_\ell^{m_2,m_3}&=&
i m_{2,3} \, \mathcal{Y}_\ell^{m_2,m_3}
\eea

The vector spherical harmonics come in three classes, which we choose to be also eigenfunctions of
$\partial_{2,3}$. We have longitudinal harmonics
$\mathcal{H}_i=\nabla_i \mathcal{Y}_\ell^{m_2,m_3}$, $\ell \geq 1$  which 
are in the $(\frac{\ell}{2}, \frac{\ell}{2})$ 
representation of $SO(4)$ with $(m_2,m_3)$ ranging as before. 
Then, there are two types of transverse vector spherical harmonics: $\mathcal{M}^{+}_i \equiv \mathcal{Y}_i^{(\ell,m_2,m_3);+}$ with $\ell\geq 1$ 
in the $\left(\frac{\ell- 1}{2},\frac{\ell+1}{2}\right)$ and 
$\mathcal{M}^{-}_i \equiv \mathcal{Y}_i^{(\ell,m_2,m_3);-}$ with $\ell\geq 1$ 
in the $\left(\frac{\ell+1}{2},\frac{\ell-1}{2}\right)$. Their degeneracy is $\ell(\ell+2)$ and it is
counted by $|m_2\pm m_3| = \ell \mp 1-2k$ for $\mathcal{M}^+_i$ and  $|m_2\pm m_3| =\ell \pm 1- 2k$ for $\mathcal{M}^-_i$ . 
In particular, these harmonics satisfy
\bea\label{proM}
\nabla_i\nabla^i \mathcal{M}^\pm_j-R^k_j \mathcal{M}^\pm_k&=&-(\ell+1)^2 \, \mathcal{M}^\pm_j\non\\
\epsilon_{ijk}\nabla^j \mathcal{M}^{\pm;k}&=& \pm \,\sqrt{\tilde{g}} \,(\ell+1)\, \mathcal{M}^\pm_i\non\\
\nabla^i\mathcal{M}^\pm_i&=&0\non\\
\partial_{2,3}\,\mathcal{M}_i^{\pm}&=&
i m_{2,3} \, \mathcal{M}_i^{\pm}
\eea
where $R^k_j=2\delta^k_j$ is the Ricci tensor on $S^3$.

\vskip 12pt
As a first application we consider the undeformed case. Setting $\hg_i=0$, solutions corresponding to a non--trivial dispersion relation ($\omega\neq 0$) satisfy
\bea\label{uneom}
&&\left(\mathcal{O}_{0}-1\right)\,\chi_k=0\qquad \mathcal{O}_{0}\,\rho + \frac{2 R (R^2-r_0^2)}{r_0}\,\partial_t \Phi=0\qquad \mathcal{O}_{0}\,\Phi-\frac{2 R r_0}{(R^2-r_0^2)}\,\partial_t \rho=0\non\\
&&\,\,\,\,\,\,\,\,\,\,\,\,\,\qquad\qquad\quad\mathcal{O}_{0}\,A_j-\nabla^i \nabla_j A_i =0\qquad\qquad\quad\,\, \nabla^i A_i=0
\eea
where now $\Phi \equiv \varphi$.
We see that the scalar and gauge fluctuations decouple. In particular, the scalar equations give the usual~\cite{DJM}
\beq\label{omeganon}
\omega^{(0)}_k=\frac{(\ell+1)}{R}\qquad\omega^{(0)}_I=\frac{(\ell+2)}{R}\qquad\omega^{(0)}_{II}=\frac{\ell}{R}
\eeq
On the other hand, expanding $A_i$ in vector spherical harmonics 
corresponds to turning off the longitudinal modes $\mathcal{H}_i$ since $\nabla^i A_i=0$ from (\ref{uneom}) and $\ell\geq 1$. We will find that in the deformed case these longitudinal modes will play a crucial role in determining the right bosonic spectrum of fluctuations. Non--vanishing undeformed solutions come from the transverse vector spherical harmonic expansions (\ref{proM}) with frequencies \beq\label{gaunon}
\omega^{(0)}_\pm=\frac{(\ell+1)}{R}
\eeq 
(see the next section and impose $\hg_i\equiv 0$ for more details). To conclude, we would like to stress that the above results accomplish the study of~\cite{DJM} for undeformed giant gravitons, taking into account the contributions of the worldvolume gauge fields~\footnote{The gauge field analysis for giants in the undeformed PP--wave scenario has been performed in~\cite{SS}.}. In particular, we observe that the frequencies (\ref{gaunon}) coincide with the four $\omega^{(0)}_k$ in (\ref{omeganon}). However, with a suitable shift in $\ell$, the undeformed spectrum is completely degenerate and we will comment on this interesting point in Section~\ref{disc}. Now we proceed to investigate the main case $\hg_i\neq 0$.

\subsection{Decoupled modes}

The scalar modes $\chi_k$ are decoupled from the rest and equation (\ref{ceom}) has been solved in~\cite{godb} expanding the perturbations as
\beq
\chi_k(t,\theta,\phi_2,\phi_3) = X_{k} \,e^{-i \omega_k t}\,\mathcal{Y}_{\ell}^{m_2,m_3}(\theta,\phi_2,\phi_3)\non
\eeq
Their frequencies are given by
\beq\label{omegak}
\omega_k^2=\frac{1}{R^2}\left[(\ell+1)^2+\hat{\Gamma}^2\right] 
\eeq
where we have defined the positive quantity 
\beq\label{gammaq}
\hat{\Gamma}^2=\left(1-\frac{r_0^2}{R^2}\right)(\hat{\gamma}_2 m_3-\hat{\gamma}_3 m_2)^2
\eeq
The fact that the four fluctuations in $AdS_5$ are completely blind to the gauge fields is an expected result since the giant graviton worldvolume, where the vector modes live, lies inside the deformed five--sphere.

\vskip 12pt
\noindent
Being in a different representation, the harmonics $\mathcal{M}^\pm_i$ do not mix with the others, so we can make the ansatz 
\beq\label{Imod}
\rho=0\qquad \Phi=0\qquad A_i(\sigma^a)=X_\pm\,e^{-i \omega_\pm t}\,\mathcal{M}^\pm_i(\theta, \phi_2,\phi_3)
\eeq
By using the identity $\nabla^i A_i=0$ as follows from (\ref{proM}), equations (\ref{reom})--(\ref{ti}) 
are identically satisfied whereas equation (\ref{ii}) now reads
\beq
\mathcal{O}_{\hg}\,A_j - \nabla^i \nabla_j A_i = 0
\eeq
Considering the explicit expression for the operator $\mathcal{O}_{\hg}$ 
in (\ref{ogamma}) and using the
properties (\ref{proM}) we find a non--trivial solution when
\beq\label{omegapm}
\omega_\pm^2=\frac{1}{R^2}\left[(\ell+1)^2+\hat{\Gamma}^2\right] 
\eeq
which is exactly the same value found in (\ref{omegak}).

Note that the role of the deformation is to add the same positive quantity $\hat{\Gamma}^2/R^2$ to all undeformed frequencies.

\subsection{Coupled vibrations}   

Finally, we consider the following coupled fluctuations
\bea\label{modesIII}
\rho(\sigma^a)&=& X_\rho\, e^{-i \omega t}\,\mathcal{Y}^{m_2,m_3}_\ell(\theta,\phi_2,\phi_3) 
\non\\
\Phi(\sigma^a)&=&X_\Phi\, e^{-i \omega t}\,\mathcal{Y}^{m_2,m_3}_\ell(\theta,\phi_2,\phi_3) 
 \\
A_i(\sigma^a)&=&X_A\, e^{-i \omega t}\,\nabla_i\mathcal{Y}^{m_2,m_3}_\ell(\theta,\phi_2,\phi_3)\non
\eea
with $\ell \geq 1$.
Inserting these expansions in (\ref{reom})--(\ref{ii}) and using the identities (\ref{Yprop}) for the scalar harmonics, after a bit of algebra
we obtain the matrix equation
\beq\label{sys}
\left[
\begin{array}{ccccc}
R^2 \omega ^2 - \ell(\ell+2)-\hat{\Gamma}^2 & $\,\quad$ & -2 i \omega\frac{R (R^2-r_0^2)}{r_0} & $\,\quad$ & 0\\
2 i \omega\frac{R r_0}{(R^2-r_0^2)}\left(1-\frac{\hat{\Gamma}^2}{R^2 \omega^2}\right) & $\,\quad$ & R^2 \omega ^2 - \ell(\ell+2)-\hat{\Gamma}^2 & $\,\quad$ & 0\\
\frac{2 r_0}{R \omega} (\hg_2 m_3-\hg_3 m_2) & $\,\quad$ & -i (R^2-r_0^2) (\hg_2 m_3-\hg_3 m_2) & $\,\quad$ & \ell(\ell+2)\\
0 & $\,\quad$ & -i (R^2-r_0^2) (\hg_2 m_3-\hg_3 m_2) & $\,\quad$ & R^2 \omega^2 - \hat{\Gamma}^2
\end{array}
\right]
\left[
\begin{array}{c} 
X_\rho \\ 
X_\Phi\\
X_A
 \end{array} 
\right]=0
\eeq
These are four equations for three unknowns 
$X_\rho,\, X_\Phi,\, X_A$ and lead to non--trivial
solutions only if they are compatible, namely if the rank $R(M)< 3$, where $M$ is the above matrix of coefficients.
This happens if 
\beq\label{omegaIII}
\omega^2_I=\frac{1}{R^2}\left[(\ell+2)^2+\hat{\Gamma}^2\right]\qquad\qquad\omega^2_{II}=\frac{1}{R^2}\left(\ell^2+\hat{\Gamma}^2\right)
\eeq
and $R(M)=2$. Again we find the same factor $\hat{\Gamma}^2$ modifying the undeformed frequencies.
Before closing this section we comment on the particular $\ell=m_2=m_3=0$ case. 
In (\ref{modesIII}) this 
corresponds to switching off $A_i$ since they turn out to be independent of the three--sphere
coordinates and of course we also have $\hat{\Gamma}^2=0$. The system (\ref{sys}) reduces to the standard undeformed one and yields the frequencies $\omega_I=2/R$ and $\omega_{II}=0$~\footnote{Recall that as done in the undeformed case~\cite{DJM}, we remove the zero modes $\omega_{II}=0$ and also $\omega_k=1/R$ in (\ref{omegak}) from the spectrum since they correspond to the collective motion of the brane and change its conserved quantum numbers. They can no longer be viewed to belong to the giant we started with.}. We now analyze in detail the degeneracies that characterize the spectrum of vibration modes around the giant graviton.

\subsection{Discussion}\label{disc}

For a giant graviton expanded in the deformed sphere part of the geometry, we have found four scalar fluctuations into $AdS_5$ (with frequencies $\omega^2_k$) and two within $\tilde{S}^5$ ($\omega^2_{I,II}$). Furthermore, it has two bosonic vibrations coming from the pure vector expansion of the worldvolume gauge fields ($\omega^2_\pm$). In particular, putting together (\ref{omegak}) and (\ref{omegapm}), and from (\ref{omegaIII}) we get
\bea\label{omega}
\omega^2_k=\omega_\pm^2=\frac{1}{R^2}\left[(\ell+1)^2+\hat{\Gamma}^2\right]\qquad\omega^2_I=\frac{1}{R^2}\left[(\ell+2)^2+\hat{\Gamma}^2\right]\qquad\omega^2_{II}=\frac{1}{R^2}\left(\ell^2+\hat{\Gamma}^2\right)\non\\
\eea
The important result we would like to stress again is that the role of the gauge field fluctuations is crucial in giving the right shape to the full bosonic spectrum. What happens is that the whole set of undeformed frequencies obtained in~\cite{DJM} together with the new (\ref{gaunon}) get modified by a universal positive shift in the present $\hg_i\neq 0$ case~\footnote{For any value of $\hg_i$ there are no tachyonic modes, so confirming the stability 
of the configuration already discussed in~\cite{godb}.}. In fact, we can rewrite the frequency of a generic vibration $X$ as 
\beq
\omega^2_X(\ell,m_2,m_3)= \frac{1}{R^2}\,\left[\left(\omega_X^{(0)}(\ell)\right)^2 +\hat{\Gamma}^2(m_2,m_3)\right]
\label{MX}
\eeq 
where $\omega_X^{(0)}(\ell)$ is the undeformed frequency (see (\ref{omeganon}) and (\ref{gaunon})), whereas $\hat{\Gamma}^2(m_2,m_3)$ defined in (\ref{gammaq}) is the universal splitting term that induces an explicit dependence on the quantum numbers $(m_2,m_3)$ with a pattern similar to the Zeeman effect for atomic electrons. A similar behavior has been found in the meson mass splitting of~\cite{mim} and the following analysis is only intended to mimic those studies on the degeneracies characterizing the spectrum. 

So, we study the general case $\ell \geq 0$ in which this initial degeneracy occurs:
\beq
\omega^{(0)}_{k,\pm}(\ell+1)=\omega^{(0)}_{I}(\ell)=\omega^{(0)}_{II}(\ell+2)
\label{shift}
\eeq
In particular, in the undeformed $\hg_i=0$ case, this allows $8(\ell+2)^2$ bosonic degrees of freedom to have the same frequency. When we turn on the deformation, a complete degeneracy exists among states that satisfy the above condition and are characterized by the same value of $\hat{\Gamma}^2(m_2,m_3)$. Therefore, having performed the $\ell$--shift as in (\ref{shift}), we concentrate on the degeneracy in $\hat{\Gamma}^2(m_2,m_3)$ for fixed values of $\ell$.

\begin{figure} [h]
\begin{center}
\epsfysize=4.0cm\epsfbox{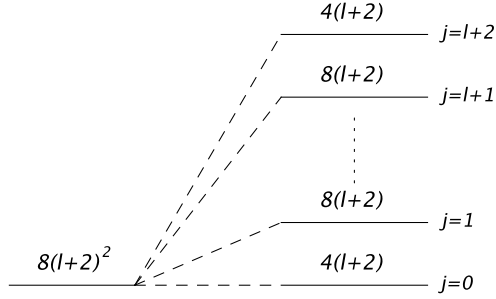}
\end{center}
\caption{The Zeeman--splitting of the undeformed $8(\ell+2)^2$ degrees of freedom for $\hg_2=\hg_3$.}
\label{susy}
\end{figure}

\vskip 12pt
In the $\hg_2=\hg_3 \equiv\hg$ case, the deformation enters the spectrum only through the difference $(m_2-m_3)$ so that the splitting term $\hat{\Gamma}^2$ depends only on a single integer $j$ as
\beq\label{sumas}
j \equiv |m_2-m_3|=0,1,\cdots,\ell+2  \qquad\quad\,\,\ \, \hat{\Gamma}^2(j) =\left(1-\frac{r_0^2}{R^2}\right)\,\hg^2\,j^2
\eeq
For any value of $\ell \geq 0$ we observe a Zeeman--like splitting as shown in Figure~\ref{susy}.
\\
Precisely, the splitting occurs in the following way:
There are $4(\ell+2)$ degrees of freedom corresponding to $j=0$ and $j=\ell+2$, and $8(\ell+2)$ for $j=1,\cdots,\ell+1$. The total number of states sum up correctly to $8(\ell+2)^2$. 

This scenario of course includes the supersymmetric LM--theory but recall that the value of $\hg_1$ never enters the game so that the non--supersymmetric case $\hg_2=\hg_3\neq\hg_1$ seems to be a special at a bosonic level. It would be very interesting to see what role the fermionic fluctuations play in this particular situation. 

\begin{figure} [h]
\begin{center}
\epsfysize=6cm\epsfbox{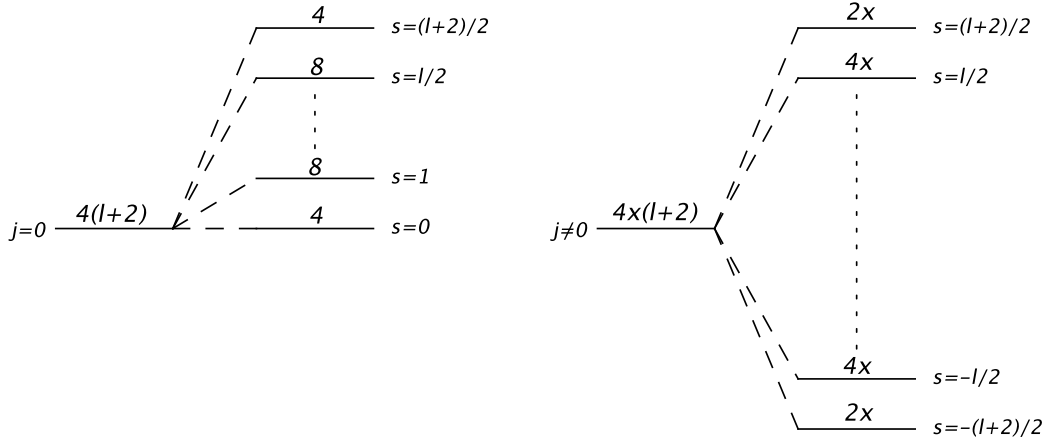}
\end{center}
\caption{The Zeeman--splitting of the $\hg_2=\hg_3=\hg$ degrees of freedom for $\hg_2\neq\hg_3$ and  $(\ell+j)$ even. We have $x=2$ for $j\neq \ell+2$ and $x=1$ for $j=\ell+2$. The value of $\hat{\Gamma}^2$ here appearing is pictured considering the case $\hg_3<\hg<\hg_2$.}
\label{nongiapar}
\end{figure}

\begin{figure} [h]
\begin{center}
\epsfysize=6cm\epsfbox{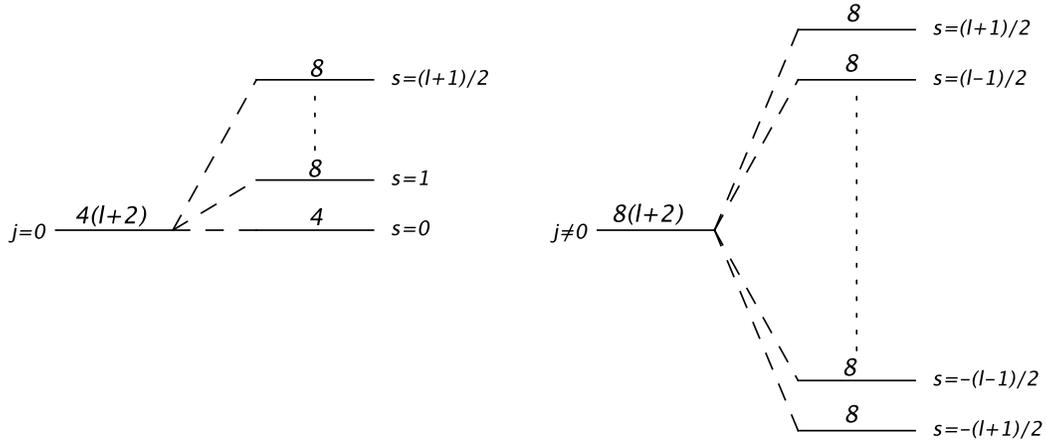}
\end{center}
\caption{The Zeeman--splitting of the $\hg_2=\hg_3$ degrees of freedom for $\hg_2\neq\hg_3$ and $(\ell+j)$ odd. Once again $\hg_3<\hg<\hg_2$.}
\label{nongiadis}
\end{figure}

\vskip 12pt 
In the more general case $\hg_2\neq\hg_3$, the splitting term $\hat{\Gamma}^2$ now depends on both $m_{2,3}$ and no longer on their difference. In order to make the comparison with the $\hg_2=\hg_3$ case easier, for fixed $\ell$ it is convenient 
to label $\hat{\Gamma}^2$  by the two numbers $j$ and $s$ 
\beq\label{nomas}
\hat{\Gamma}^2(j,s) = \left(1-\frac{r_0^2}{R^2}\right)\,\left[\left(\frac{j}{2}+s\right)\,\hg_2+\left(\frac{j}{2}-s\right)\,\hg_3\right]^2
\eeq
where $j$ is still defined as in (\ref{sumas}), while $s$ takes different values according to $(\ell+j)$ being even or odd. Its range reads
\bea
&&\mbox{$(\ell+j)\qquad$ even}\qquad\qquad
\left\{
\begin{array}{l}
s=0,\cdots,\frac{\ell+2}{2}\qquad\quad\,\,\,\, j=0\\
s=-\frac{\ell+2}{2},\cdots,\frac{\ell+2}{2} \qquad j\neq 0
    \end{array}
\right.\\
\non\\
&&\mbox{$(\ell+j)\qquad$ odd}\qquad\qquad\,\,
\left\{
\begin{array}{l}
s=0,\cdots,\frac{\ell+1}{2}\qquad\quad\,\,\,\, j=0\\
s=-\frac{\ell+1}{2},\cdots,\frac{\ell+1}{2} \qquad j\neq 0
    \end{array}
\right.
\eea
By fixing $j$, the degenerate degrees of freedom of the $\hg_2=\hg_3$ case further 
split according to the different values of $s$.  
If $(\ell+j)$ is even and $j=0$, the previous $4(\ell+2)$ degenerate levels split in $(\ell/2+2)$ new levels while for $j\neq 0$ the $4x(\ell+2)$ levels open up in $(\ell+3)$ levels where $x=2$ for $j\neq \ell+2$ and $x=1$ for $j=\ell+2$ (see Figure \ref{nongiapar}).\\
\\
On the other hand, a more symmetric splitting occurs if $(\ell+j)$ is odd. In fact, for $j=0$ we 
find $(\ell+3)/2$ different frequency levels and when $j\neq 0$ they are $(\ell+2)$ as depicted in Figure \ref{nongiadis}. 

\section{Dual giant gravitons}\label{dggf}

The analysis just performed signals the same $\hg_i$--behavior for all the fluctuations around a giant expanded in the deformed part of the background. What about the dual giant case?
The answer follows from revising the fluctuations around the dual giant configuration found in~\cite{godb}, now extended in order to understand the worldvolume gauge field contributions. Recall that the stable D3--brane  has constant radius ($l_0$), orbits rigidly in the $\phi_1$ direction on the deformed $\tilde{S}^5$ and is wrapped in the three--sphere $(\alpha_1,\alpha_2,\alpha_3)$ contained in the $AdS_5$ part of the geometry
\beq
ds^2_{AdS_5}=-(1+\frac{l^2}{R^2})dt^2+\frac{dl^2}{1+\frac{l^2}{R^2}}+l^2\left[d\alpha_1^2+\sin^2{\alpha_1}\left(d\alpha_2^2+\sin^2{\alpha_2}d\alpha_3^2\right)\right]
\eeq
now written in terms of global coordinates as in (\ref{old}).
Again, we study the role played by the gauge field fluctuations on the worldvolume of the probe. We take $F\neq 0$,
\beq\label{an}
l=l_0+\varepsilon\,\lambda(\sigma^a)\,\qquad \phi_1=\omega_0 t+\varepsilon\,\varphi_1(\sigma^a)
\eeq
and
\beq\label{ans}
r=\varepsilon\,\rho(\sigma^a)\,\qquad \theta=\frac{\pi}{4}+\varepsilon\,\vartheta(\sigma^a)\,\qquad \phi_2=\varepsilon\,\varphi_2(\sigma^a)\,\qquad \phi_3=\varepsilon\,\varphi_3(\sigma^a)
\eeq
with $\sigma^a=(t,\alpha_1,\alpha_2,\alpha_3)$. We are simply perturbing the equilibrium dual giant configuration reviewed in Section~\ref{rev}, see (\ref{ansadu}), here recovered by setting $\varepsilon=0$ in (\ref{an}) and (\ref{ans}). Since the calculation runs parallel to that of the previous section we are free to omit the details and we only report the final form of the equations of motion. For the scalars we have
\beq\label{eomlam}
\partial_a\left(\sqrt{\tilde{g}}\,\mathcal{G}^{ab}\,\partial_b\l \right)+\frac{2 R (R^2+l_0^2)}{l_0}\,\sqrt{\tilde{g}}\,\partial_t \varphi_1=0
\eeq
\beq\label{eomphid}
\partial_a\left(\sqrt{\tilde{g}}\,\mathcal{G}^{ab}\,\partial_b\varphi_1 \right)-\frac{2 R l_0}{(R^2+l_0^2)}\, \sqrt{\tilde{g}}\,\partial_t \l=0
\eeq
\beq\label{eomrhod}
\partial_a\left(\sqrt{\tilde{g}}\,\mathcal{G}^{ab}\,\partial_b\rho\right)-\sqrt{\tilde{g}}\left(1+\tilde{\Gamma}^2\right)\,\rho=0
\eeq
while the equations of motion for the gauge fields, coming from their second order Lagrangian, take the simple form
\bea\label{eomad}
\mathcal{L}^{(2)}_{gauge}=-\frac{T_3}{4 R}\,\sqrt{\tilde{g}}\,\left( \mathcal{G}^{ac}\mathcal{G}^{bd} F_{ab} F_{cd}\right)\quad\Rightarrow\quad\partial_a\left(\sqrt{\tilde{g}}\,\mathcal{G}^{ac}\mathcal{G}^{bd}\,F_{cd}\right)=0
\eea
They are decoupled from the scalar modes and do not depend on the deformation parameters since the dual giant worldvolume lies in the undeformed part of the geometry. The matrix $\mathcal{G}$ is the same as in (\ref{unmat}) but now $\tilde{g}_{ij}$ is the diagonal metric on the unit three--sphere $(\alpha_1,\alpha_2,\alpha_3)$ with determinant $\tilde{g} = \sin^4{\alpha_1} \sin^2{\alpha_2}$. On the other hand, the positive quantity $\tilde{\Gamma}^2$ reads
\beq\label{tilga}
\tilde{\Gamma}^2=\left(1+\frac{l_0^2}{R^2}\right)\,\left(\frac{\hg_2^2+\hg_3^2}{2}\right)
\eeq 
The first three equations (\ref{eomlam})--(\ref{eomrhod}) are exactly the same found in~\cite{godb} and give 
\beq\label{omegar}
\tilde{\omega}^2_\rho=\frac{1}{R^2}\left[(\ell+1)^2+\tilde{\Gamma}^2\right]\qquad\tilde{\omega}^2_I=\frac{(\ell+2)^2}{R^2}\qquad\tilde{\omega}^2_{II}=\frac{\ell^2}{R^2}
\eeq
Rewriting the metric of the deformed five--sphere (\ref{sdef}) using the complete set of coordinates $\{\phi_1,y_k\}$, with $k=1,\cdots,4$, as done in~\cite{DJM}, we can easily see that the radial frequency $\tilde{\omega}_\rho^2$ in (\ref{omegar}) corresponds to the four $\tilde{\omega}^2_k$ as~\footnote{We have checked this explicitly. However it is possible to coherently consider $\tilde{\omega}_\rho$ taking into account that its degeneracy is equal to $4$.}
\beq\label{omegacappa}
\tilde{\omega}^2_\rho\,\longrightarrow\,\tilde{\omega}^2_k=\frac{1}{R^2}\left[(\ell+1)^2+\tilde{\Gamma}^2\right]
\eeq
Equations (\ref{eomad}) still come in two classes. With the gauge choice $A_t=0$, we also have $\nabla^i A_i =0$ and a non--trivial solution exists for mode expansions involving the vector spherical harmonics $\mathcal{M}^\pm_i$ (\ref{proM}), exactly as in the undeformed giant graviton case discussed in Section~\ref{vib}. The resulting frequencies are
\bea\label{omegapma}
\tilde{\omega}_{\pm}^2&=&\frac{(\ell+1)^2}{R^2}
\eea
which do not depend on the deformation parameters. Still, we can perform a suitable shift in $\ell$ to study degeneracies. In the undeformed case we have $\tilde{\Gamma}^2=0$ and
 \beq
\tilde{\omega}_{k,\pm}(\ell+1)=\tilde{\omega}_{I}(\ell)=\tilde{\omega}_{II}(\ell+2)
\eeq
so that the spectrum can be characterized by the same frequencies. On the other hand, if $\hg_i\neq 0$ the degeneracy in $\ell$ is partially broken by the non--vanishing $\tilde{\Gamma}^2$ which increases the contributions of the four transverse fluctuations of the dual giant inside the deformed $\tilde{S}^5$, see (\ref{omegacappa}). In particular, we have that the original $8(\ell+2)^2$ undeformed degenerate degrees of freedom are now split into two sets. There are $4(\ell+2)^2$ with the same undeformed frequency $\omega\sim (\ell+2)$, while the remaining $4(\ell+2)^2$ encode the dependence on the deformation parameters through the shift--quantity (\ref{tilga}).

\section{Conclusions}

In this paper we have provided an extension of the results found in~\cite{godb} for the fluctuations around ground state configurations of (dual) giant gravitons in Lunin--Maldacena and Frolov
backgrounds. Since worldvolume gauge fluctuations are also computed, our knowledge of their full bosonic spectrum is now complete. Both the ${\cal N}=1$ and non--supersymmetric cases, depending on the choice of the deformation 
parameters $\hat{\g}_i$, have been analyzed.

When the giant lies in the deformed part of the geometry, we have found that a non--trivial dependence on the $\hat{\gamma}_{2,3}$ parameters
appears both in terms that correct the free dynamics of the modes and in terms that couple
the $U(1)$ worldvolume gauge field to the
scalars in the orthogonal directions to the dynamical D3--brane.
The $\hat{\gamma}_1$ parameter, associated with a TsT transformation along the torus inside 
the D3 worldvolume, never enters the equations of motion. The situation is closely related to the one found in~\cite{mim} and it could be an expected result. In fact, in both cases, the branes wrap the same deformed sphere in the internal space, but we would like to stress that the giant graviton is dynamical whereas the flavor D7 is a spacetime filling brane. The geometry of the configurations, but not their dynamical properties, seems to rule the general $\hg_i$--behavior in the quadratic fluctuations around these D--brane ground states.    

A smooth limit to the undeformed equations of motion exists
for $\hat{\gamma}_i \rightarrow 0$. 
In this limit all the modes decouple and we are back to the undeformed solutions of~\cite{DJM}, now extended with the gauge field contributions. This is a nice result, which easily follows from our more involved analysis. In fact, the situation heavily changes once we consider
the general deformed equations with $\hg_i\neq 0$. Analytically solving these equations for
elementary excitations of scalars and vectors, we have found that  
the spectrum is slightly modified with respect to the one found in~\cite{godb}. The new deformed frequencies acquire a non--trivial dependence on $\hat{\gamma}_{2,3}$ that is universal: The whole set of undeformed frequencies gets shifted by the same quantity and the inclusion of gauge fluctuations in the game is the nodal point. 
The universal term, being proportional to the $U(1) \times U(1)$ quantum numbers $(m_2,m_3)$,
induces a level splitting in a Zeeman--like effect, which strongly resembles the one observed in~\cite{mim} along the study of the mesonic spectrum in flavored marginally deformed AdS/CFT.
We have performed a detailed analysis of the level splitting and of the corresponding
degeneracy.
The situation turns out to be very different depending on whether $\hg_2$ and $\hg_3$ are equal or not. For $\hg_2 \neq \hg_3$ the degeneracy is almost completely broken, while for $\hg_2=\hg_3$   
the frequency levels are characterized by a weak splitting effect.
However, as already observed in~\cite{godb}, the spectrum is deformed by a quantity whose strength depends unavoidably on the radius of the giant: The larger the giant, the smaller the deformation splitting. 

By studying giant gravitons on a deformed $(J,0,0)$ PP--wave, the authors of~\cite{HM} also found a classical configuration independent of the deformation and with a spectrum of small fluctuations almost identical to the one obtained in~\cite{godb}. Since they do not consider possible couplings among scalar and vector modes, the universal $\hg$--behavior we find could also appear upon turning on the gauge fields on the worldvolume of the PP--wave giants and change the precise value of their frequencies. It would be very interesting to check this expectation~\footnote{Another work dealing with giant gravitons on deformed PP--wave background is~\cite{gre}.}.

The scenario is totally different for dual giant gravitons. In Section~\ref{dggf} we have also introduced gauge field fluctuations on the worldvolume of a dynamical D3--brane lying in AdS. Scalar and vector modes are decoupled since only the deformation parameters drive their mixing and the worldvolume of the brane is now inside $AdS_5$ which is not affected by the deformation. Only the vibrations of the dual in the the deformed five--sphere directions depend on $\hg_{2,3}$ together with the radius of the compact brane. The remaining scalar fluctuations coincide with the undeformed ones; for the gauge fields the deformation is also harmless. This provides the full analysis of the spectrum around dual giants even in the standard undeformed $AdS_5\times S^5$ background, so enhancing the studies of~\cite{DJM}. 

Let us conclude by mentioning some directions in which our work could be extended. First, it is very important to reproduce the spectrum of fluctuations about these giant configurations and find their dual CFT operator counterparts. Furthermore, to fulfill the deformed spectrum around (dual) giants and learn about its supersymmetry properties one should also study the fermionic sector.  In fact, restricting to the LM $\mathcal{N}=1$ case, our giant gravitons should preserve supersymmetry since they saturate a BPS bound (see the approach proposed in~\cite{marioni} and~\cite{forbu} where, in particular, the last analyzes in details the supersymmetric dual giant case) and we expect to see supersymmetry in action in a suitable reorganization of the bosonic/fermionic spectra. We leave these and other interesting open problems for the future.

\vspace{0.1cm}

\section*{Acknowledgments}

\noindent 
I would like to thank Alberto Mariotti, Sean McReynolds, Alberto Zaffaroni and especially Silvia Penati for fruitful discussions and comments. This work has been supported in part by INFN, PRIN prot. $2005-024045-004$ and the European Commission RTN program MRTN--CT--2004--005104.

\newpage



\begin{thebibliography}{99}
\parskip-2pt

\bibitem{godb}
  M.~Pirrone,
  JHEP {\bf 0612}, 064 (2006)
  [arXiv:hep-th/0609173].





\bibitem{LM}
O.~Lunin and J.~Maldacena,
JHEP {\bf 0505} (2005) 033 [arXiv:hep-th/0502086].


\bibitem{F}
S.~Frolov,
JHEP {\bf 0505}, 069 (2005)
[arXiv:hep-th/0503201].


\bibitem{MGST}
J.~McGreevy, L.~Susskind and N.~Toumbas,
JHEP {\bf 0006}, 008 (2000)
[arXiv:hep-th/0003075].


\bibitem{GMT}
M.~T.~Grisaru, R.~C.~Myers and O.~Tafjord,
JHEP {\bf 0008}, 040 (2000)
[arXiv:hep-th/0008015].


\bibitem{HHI}
A.~Hashimoto, S.~Hirano and N.~Itzhaki,
JHEP {\bf 0008}, 051 (2000)
[arXiv:hep-th/0008016].


\bibitem{mim}
  S.~Penati, M.~Pirrone and C.~Ratti,
  arXiv:0710.4292 [hep-th].

\bibitem{DJM}
S.~R.~Das, A.~Jevicki and S.~D.~Mathur,
Phys.\ Rev.\ D {\bf 63}, 024013 (2001)
[arXiv:hep-th/0009019].



\bibitem{my}
  M.~Kruczenski, D.~Mateos, R.~C.~Myers and D.~J.~Winters,
  JHEP {\bf 0307}, 049 (2003)
  [arXiv:hep-th/0304032].



\bibitem{SS}
  D.~Sadri and M.~M.~Sheikh-Jabbari,
  Nucl.\ Phys.\  B {\bf 687}, 161 (2004)
  [arXiv:hep-th/0312155].




\bibitem{HM}
  A.~Hamilton and J.~Murugan,
  JHEP {\bf 0706}, 036 (2007)
  [arXiv:hep-th/0609135].


\bibitem{gre}
  S.~D.~Avramis, K.~Sfetsos and D.~Zoakos,
  Nucl.\ Phys.\  B {\bf 787}, 55 (2007)
  [arXiv:0704.2067 [hep-th]].

\bibitem{marioni}
  A.~Mariotti,
  JHEP {\bf 0709}, 123 (2007)
  [arXiv:0705.2563 [hep-th]].

\bibitem{forbu}
  A.~Butti, D.~Forcella, L.~Martucci, R.~Minasian, M.~Petrini and A.~Zaffaroni,
  arXiv:0712.1215 [hep-th].


\end{thebibliography}
\end{document}